\documentclass[conference]{IEEEtran}
\IEEEoverridecommandlockouts
\usepackage{cite}
\usepackage{amsmath,amssymb,amsfonts}
\usepackage{algorithmic}
\usepackage{graphicx}
\usepackage{multirow}
\usepackage{makecell}
\usepackage{textcomp}
\usepackage{enumitem}
\usepackage{siunitx}
\usepackage{caption}
\usepackage{titlesec}
\usepackage{xcolor,colortbl}

\def\BibTeX{{\rm B\kern-.05em{\sc i\kern-.025em b}\kern-.08em
    T\kern-.1667em\lower.7ex\hbox{E}\kern-.125emX}}

\usepackage[all]{background}
\usepackage{stackengine}
\setstackEOL{\\}
\setstackgap{L}{\normalbaselineskip}
\SetBgContents{\color{gray}{\tiny \Longstack{PREPRINT - To appear at VLSI-SoC 2024}}}
\SetBgPosition{4.5cm,1cm}
\SetBgOpacity{1.0}
\SetBgAngle{0}
\SetBgScale{1.8}

\begin{document}
\bstctlcite{IEEEexample:BSTcontrol}



\setlength\abovecaptionskip{0\baselineskip}
\setlength\belowcaptionskip{-1.4\baselineskip}
 
\def\@IEEEfigurecaptionsepspace{\vskip\abovecaptionskip\relax}%
\def\@IEEEtablecaptionsepspace{\vskip\abovecaptionskip\relax}%

\titlespacing*{\section}
{0pt}{-.1mm}{-.1mm}

\title{In-Memory Mirroring: Cloning Without Reading \vspace{-0.4cm}}

\author{Simranjeet Singh\IEEEauthorrefmark{1}, Ankit Bende\IEEEauthorrefmark{3}, Chandan Kumar Jha\IEEEauthorrefmark{4}, Vikas Rana\IEEEauthorrefmark{3}, \\ Rolf Drechsler\IEEEauthorrefmark{4}\IEEEauthorrefmark{5},  Sachin Patkar\IEEEauthorrefmark{1}, Farhad Merchant\IEEEauthorrefmark{2} \\
\IEEEauthorrefmark{1}IIT Bombay, India, \IEEEauthorrefmark{3}Forschungszentrum Jülich GmbH, Germany,  \\ \IEEEauthorrefmark{4}University of Bremen, Germany, \IEEEauthorrefmark{5}DFKI GmbH, Germany, \IEEEauthorrefmark{2}Newcastle University, UK \\ 
\{simranjeet, patkar\}@ee.iitb.ac.in, \{a.bende, v.rana\}@fz-juelich.de, \\ \{chajha, drechsler\}@uni-bremen.de, farhad.merchant@newcastle.ac.uk
\vspace{-0.4cm}}

\maketitle

\begin{abstract}

In-memory computing (IMC) has gained significant attention recently as it attempts to reduce the impact of memory bottlenecks. Numerous schemes for digital IMC are presented in the literature, focusing on logic operations. Often, an application's description has data dependencies that must be resolved. Contemporary IMC architectures perform read followed by write operations for this purpose, which results in performance and energy penalties. To solve this fundamental problem, this paper presents in-memory mirroring (IMM). IMM eliminates the need for read and write-back steps, thus avoiding energy and performance penalties. Instead, we perform data movement within memory, involving row-wise and column-wise data transfers. Additionally, the IMM scheme enables parallel cloning of entire row (word) with a complexity of $\mathcal{O}(1)$. Moreover, we analyzed the energy consumption of the proposed technique on an RRAM crossbar with an experimentally validated JART VCM v1b model. The IMM increases energy efficiency and shows 2$\times$ performance improvement compared to conventional data movement methods.

\end{abstract}


\section{Introduction}

The gap between the processing unit and memory leads to speed limitations known as the memory wall. This challenge is addressed by processing data within the memory and has emerged as a solution to alleviate memory bottleneck issues~\cite{Sebastian2020nature}. One solution is to design the logic-in-memory (LiM). The fundamental approach in LiM involves storing input logical states in memory cells, with the computed output remaining in memory as a logical state. Various memory technologies, including resistive random access memory (RRAM), phase change memory (PCM), and spin-transfer torque magnetic random access memory (STT-RAM), have been utilized to implement LiM.

\begin{figure}[t]
    \centering
    \includegraphics[width=\linewidth]{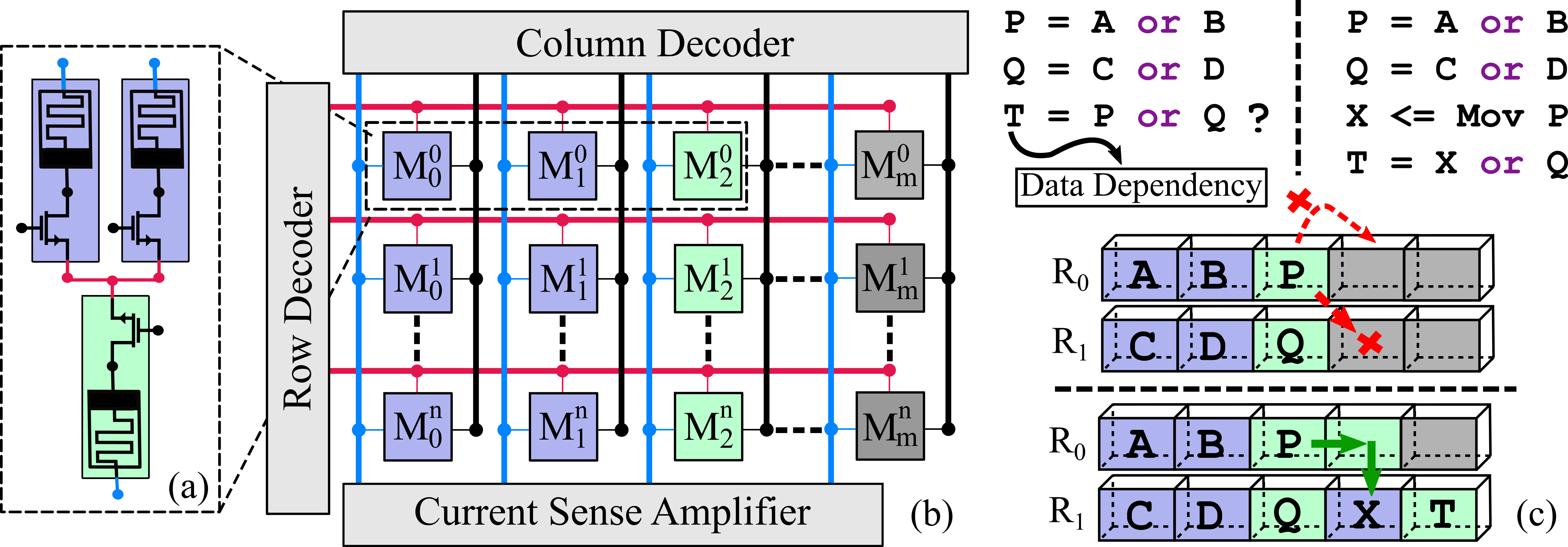}
    \caption{(a) schematic for implementing the OR operation within the crossbar architecture, utilizing three memristors. Two memristors are designated as inputs, while the third is output storage. (b) The crossbar architecture demonstrates the parallel execution of the same operation. (c) The challenge of data dependency and proposes a solution by transferring data between devices within rows and/or columns.}
    \label{fig:intro}
\end{figure}

Among these technologies, RRAM stands out as a contender for LiM, where logical states are represented by the resistive state of the device, enabling computation within the memory~\cite{waser2009}. Several schemes of LiM using RRAM devices have been demonstrated, such as MAGIC~\cite{KBL+:2014}, IMPLY~\cite{kvatinsky2013memristor}, FELIX~\cite{gupta2018felix}, and Majority~\cite{DDH+:2023}, and so on~\cite{singh2023optimize}. Fig.~\ref{fig:intro}(a) shows the schematic to implement the MAGIC OR gate in the crossbar given in Fig.~\ref{fig:intro}(b). Furthermore, experimental demonstrations of some schemes have been conducted, validating the schemes~\cite{hoffer2020experimental,bende2024experimental, Padberg2024Experimental}.  


Various architectures have been proposed to map the logic on the crossbar to compute serially and in parallel. Considering the area and latency constraint, parallel mapping has been proposed, such as CONTRA~\cite{bhattacharjee2020contra} and SIMPLE~\cite{ben2017simple}. These schemes allow the mapping of the logic function in parallel, where both the rows are columns considered. However, data dependencies arise in many cases, particularly when 
the operations are performed in multiple columns. Fig.~\ref{fig:intro}(c) presents a scenario where three operations need to be performed, with operation T dependent on data from P and Q (left code snippet). Currently, these dependencies are handled by copying the data from the current cell to the required cell, which either requires a read and write-back cycle~\cite{ben2017simple} or two complement operations~\cite{Perach_2024}, thereby resulting in increasing the overall energy and latency of the computation. 

This paper proposes an in-memory mirroring (IMM) technique to mitigate data dependency by facilitating data cloning within the memory, thereby eliminating the need for energy-intensive read and write-back operations. Unlike previous approaches, inspired by the RowClone methods for DRAM~\cite{Seshadri2013rowclone}, IMM operates directly within the crossbar memory and does not require a read cycle, improving the overall computation latency. Furthermore, IMM introduces the capability for parallel cloning of entire rows with a one-cycle latency, further enhancing its efficiency and scalability. To the best of our knowledge, our paper represents the first pioneering demonstration of IMM using an experimentally validated device model. The following are the contributions of this paper:


 
\begin{itemize}
    \item Concept of cloning: bit cloning and word cloning.
    \item Simulation analysis and validation using JART VCM v1b~\cite{Bengel2020} SPICE model aims to demonstrate the cloning data in rows, columns, and in parallel. 
    \item Finally, energy and latency calculations during the cloning operation are conducted, and the results are compared with those found in the literature.
\end{itemize}


\section{Background and Related Work}
\label{backnrw}
\subsection{Memristive Devices}
Memristive devices have emerged as a significant advancement in non-volatile memory technology. Initially proposed as a concept by Leon Chua in 1971~\cite{Chua:1971}, memristive devices have gained prominence due to their unique ability to store data by modulating resistance states~\cite{Strukov:2008}. One such device is RRAM, where resistance modulation is achieved by applying a voltage across these device terminals. In response, the resistance of the devices changes based on the magnitude and direction of the current flow. Typically, these memristive devices can be interconnected to form a crossbar structure. However, issues related to forming and sneak-path currents can arise when individual memristive devices are connected without a CMOS transistor in series. To mitigate these concerns, memristive devices are fabricated with a CMOS transistor in series, resulting in what is known as a 1T1R cell.




A memristive 1T1R cell possesses at least two distinct states: high resistive state (HRS) and low resistive state (LRS) that are mapped to Boolean logic ‘0’ and ‘1’ for LiM implementation. To simulate the characteristics of RRAM cells, several models have been introduced in the literature for characterization at the SPICE level~\cite {Staudigl:2022}. Among all the models, JART VCM v1b is particularly noteworthy as an open-source model that is based on experimental data from devices fabricated at Forschungszentrum Jülich GmbH, Germany~\cite{Bengel2020}. Moreover, the logic gate using the MAGIC design style has been experimentally validated, resembling the device from the JART VCM 1b model~\cite{bende2024experimental}. So, in this study, JART VCM v1b has been used to conceptualize and conduct simulation analyses.

\subsection{Related Work}
Previous attempts to handle the energy and latency during data dependency for LiM have been focused on synthesizing the logic function to reduce the number of copy operations~\cite {ben2017simple}. However, the problem remains the same: computation still requires some copy operations that are expensive in energy and latency. Another approach that has been recently used is performing the two-time complement operation to copy within the memory~\cite{Perach_2024}. Even though this scheme still does not require a read cycle, the copy operation needs two cycles, leading to latency and energy inefficiency. Another approach for copying using the IMPLY logic style has been demonstrated with passive devices, requiring an extra $R_{s}$ resistor and an additional isolation voltage. Adding $R_{s}$ and addressing the sneak path problem in the passive crossbar makes this method impractical in real crossbar implementations~\cite{Lei2015fastboolean}.

Although previous literature provides limited evidence of copy operations, typically performed using four devices in series~\cite{luo2024copy} or a passive devices, a comprehensive analysis of this approach is lacking. To the best of our knowledge, our paper represents the first pioneering demonstration of IMM using an experimentally validated device model.

\section{Operating Principle of Cloning}
\label{Principle}

\begin{figure}[!t]
  \begin{minipage}[!t]{0.48\linewidth}
    \centering
    \includegraphics[width=\linewidth]{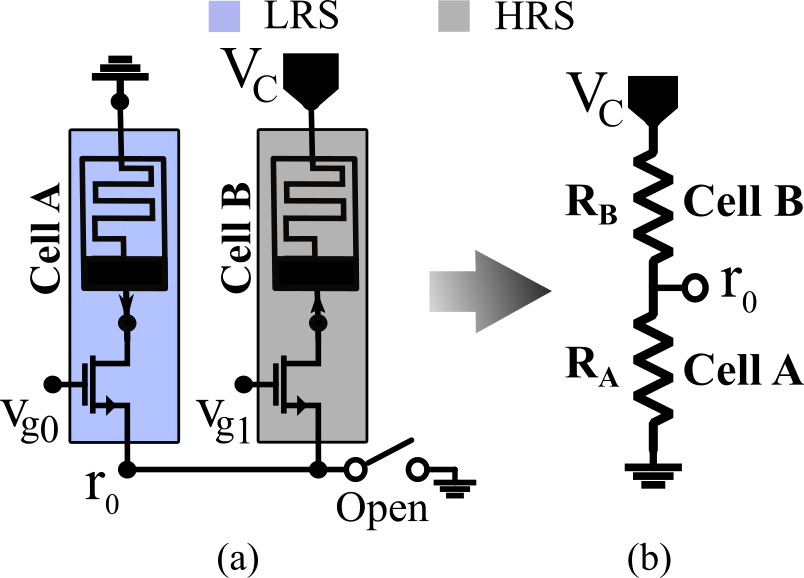}
  \caption{(a) RRAM schematic for cloning. (b) the approximate equivalent circuit. }%
  \label{fig:clone_explain}
  \end{minipage}
  \begin{minipage}[!t]{0.48\linewidth}
    \centering
\captionof{table}{Parameters (JART)}
    \label{tab:params}
    \setlength{\tabcolsep}{1.5pt}
 \begin{tabular}{|c | c|}
 \hline
     Params & Value  \\
     \hline
     \hline
    LRS ($R_{On}$) & $\approx 3.5-4.5 K\si{\ohm} $ \\
    HRS ($R_{Off}$) & $\approx 65-70 K\si{\ohm} $\\ 
    $V_{Set}$ & 1V \\
    $V_{Reset} $& $>$2.0V  \\
    $V_{Read}$ & 0.5V  \\
    $V_{C}$ & 1.5V  \\
     $V_{g}$ & 2.5V  \\
     
     \hline
\end{tabular} 
   
  \end{minipage}
\end{figure}

    



\begin{figure*}[t!]
    \centering
    \includegraphics[width=\textwidth]{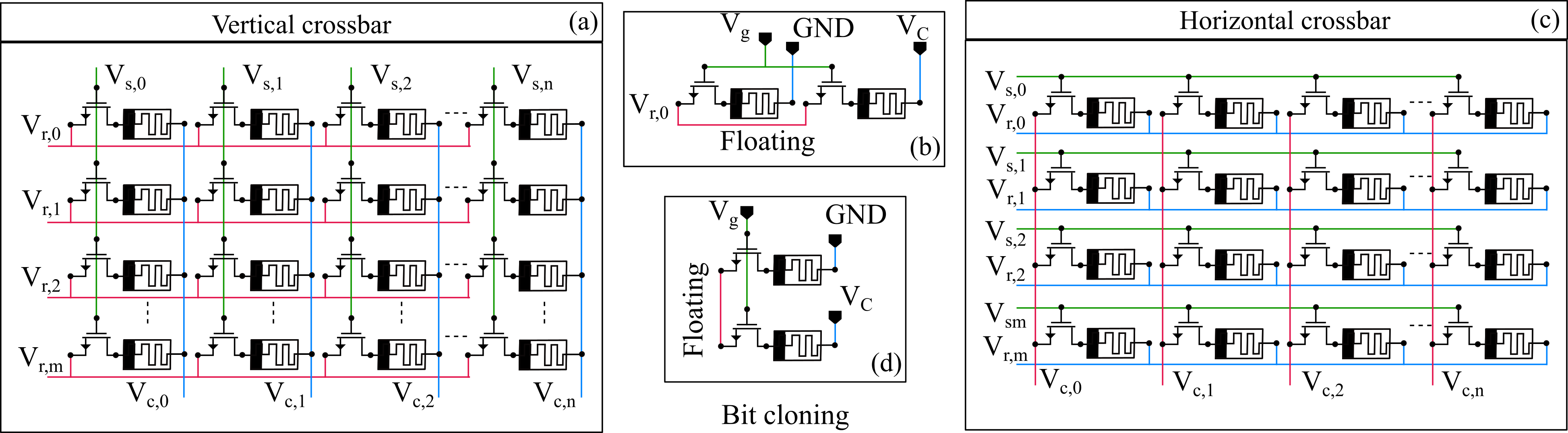}
    \caption{Different configurations of crossbar architectures. In (a), the vertical crossbar layout is presented, wherein the gates of transistors are connected vertically. (b) Bit-cloning in the vertical crossbar. (c) Horizontal crossbar, characterized by horizontally connected transistor gates. (d) Bit-cloning in horizontal crossbar}
    \label{fig:crossbars}
\end{figure*}
The IMM process utilizes just two memristors for cloning; one holds the data to be cloned, while the second is the target for cloning. Initially, all devices are considered to be in the HRS state. Fig.~\ref{fig:clone_explain}(a) shows the schematic for one-bit cloning, where two cells named Cell A and Cell B have been used. Cell A contains the data to be cloned into Cell B, which begins in the HRS state. An equivalent representation of Fig.~\ref{fig:clone_explain}(a) is shown in Fig~\ref{fig:clone_explain}(b) using the simple resistors. The resistance of Cell A is marked as $R_A$; similarly, the resistance of Cell B is marked as $R_B$. Two possible values of $R_A$ and $R_B$ could be either HRS or LRS.


In the cloning operation, a positive voltage $V_{C}$ is applied across Cell B, while Cell A is connected to the ground. Throughout this process, the gates of both transistors are fully open, and devices sharing the common line are kept open ($r_0$). This establishes a pathway between Cells A and B, where $R_{A}$ and $R_B$ are connected in series. The parameters list used for the analysis is shown in Table~\ref{tab:params}.  This creates a potential drop at $r_0$, which can be calculated according to the Equation~\ref{eq}. The cloning voltage ($V_{C}$) is chosen to be slightly greater than the Set threshold voltage ($V_{Set}$) of the memristive cell. In this case, 1.5V as $V_C$ has been chosen for a successful cloning operation. 
\begin{equation}
\label{eq}
    V_{r0} = \frac{R_A}{(R_A + R_B)} \times V_C
\end{equation}

Based on the values of $R_A$ and  $R_B$, there are two scenarios to consider: (1) when the input data ($R_A$) is logic `0' (HRS), and (2) when the input data ($R_A$) is in LRS or logic `1'. The IMM technique allows for a change of the state of $R_B$ according to the state of $R_A$.

\subsection{Case 1: Cloning Logic `1'}
In this case, Cell A contains logic `1', which needs to be cloned in Cell B and is initialized to logic `0' (HRS).
When the $V_C$ is applied, the $V_{r0}$, according to Equation~\ref{eq}, becomes very less because the $R_A << R_B$. As the $(V_C-V_{r0}) \approx V_C$ is greater than the Set voltage ($>V_{Set}$), it generates a sufficient voltage across Cell B, enabling it to transition from HRS to LRS. Cell B will switch from HRS to LRS in accordance with the data in Cell A, eventually cloning the Cell A data into Cell B. Due to the high reset-to-set ratio ($\approx 2$) in this model, Cell A will retain its original state.


\subsection{Case 2: Cloning Logic `0'}
In this scenario, Cell A is in the logic `0', and the target cell (Cell B) is already in the HRS state. According to Equation~\ref{eq}, the voltage across the row will be evenly distributed ($R_A = R_B$), and voltage at $r_0$ will be $\approx V_{C}/2$. The voltage across Cell B, $(V_C-V_{r0}) \approx V_C/2$ is not sufficient to switch the state ($<V_{Set}$). Thus, Cell B will maintain its original state of HRS,  which is the same as the state of Cell A.

\section {The Proposed Methodology}
\label{methods}
RRAM often adopts a crossbar structure, which is known for its ability to facilitate dense memory. In its simplest form, the array consists of horizontal and vertical lines, where each RRAM cell is connected at each junction. There are multiple ways to connect the RRAM cell at the junction. Fig.~\ref{fig:crossbars} shows the vertical and horizontal crossbar structure for $m \times n$ (rows $\times$ columns) size. In a vertical crossbar structure, as shown in~\ref{fig:crossbars}(a), the vertical lines are connected to the electrode of the device, and horizontal lines are connected to the transistor source. The gate of all the transistors is connected horizontally. On the other hand, in the horizontal crossbar, as shown in Fig.~\ref{fig:crossbars}(d), horizontal lines are connected to the electrode of the RRAM cell while the vertical lines are connected to a source of the transistors. Moreover, the gate of transistors is shorted horizontally. An appropriate voltage at horizontal and vertical lines is applied to write (0 or 1) and read an individual cell in the crossbar. In Fig.~\ref{fig:crossbars}, $V_{r,x}$ and $V_{c,y}$ represent the voltage at row and column lines, respectively, where $\rm |0\le x<m|$ and $|0\le y<n|$. $V_{s,z}$ is the gate switch voltage, which is given as $|0\le z<n|$ for vertical and $|0\le z<m|$ for horizontal crossbar.

In this work, the data is copied from one cell to another without performing the read operation called cloning or mirroring. Initialization of both input and output follows a standard memory write operation procedure, and the data cloning process aligns closely with this write operation method.
To integrate a cloning operation within a crossbar array, two requirements must be fulfilled: the structure of the crossbar and the connections of the memristive cells configured within the array as shown in ~\ref{fig:crossbars}. Additionally, the logical state of the memristive cell should be represented as resistance. Fig.~\ref{fig:crossbars} shows the memristive crossbar structure with the necessary connections. While the crossbar architecture allows for various other connection configurations, this study focuses on demonstrating cloning within the structures depicted in both Fig.~\ref{fig:crossbars}(a) and Fig.~\ref{fig:crossbars}(d).

\begin{figure}
    \centering
    \includegraphics[width=\linewidth]{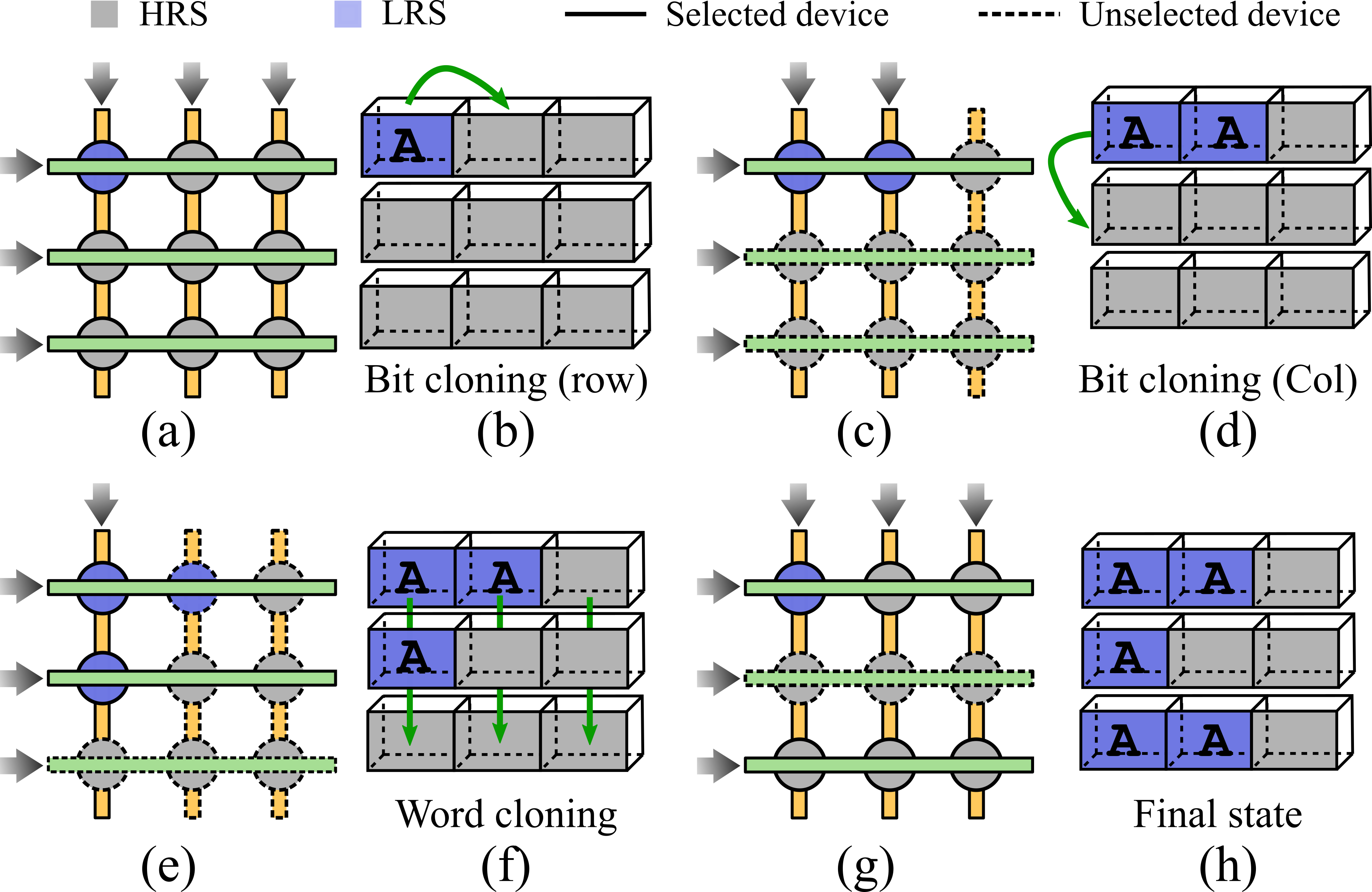}
    \caption{(a) 3x3 crossbar array structure sketch. The gray arrow exemplifies the voltage source. The colored circle at the junction represents different memristors' states. (b) Method of bit cloning in the same row, where the data is marked as ``A," which will be moved according to the representation of the green arrow. (c) Row-wise bit operations and the dotted blocks show the unselected rows and columns. (d) Column-wise bit operation where only one-bit value will be moved is marked in the green arrow. (e) Selected voltage source and devices to perform the column-wise bit cloning. (f) Representation of full column movement where the first complete row word will move to the third row in parallel.  (g) Selection of the required cell to perform word cloning. (h) final state after performing all operations from (a) to (g) in a sequence}
    \label{fig:methods}
\end{figure}

The vertical crossbar structure has a connection that is similar to the connection depicted in Fig.~\ref{fig:clone_explain}(a) for bit cloning. The $V_C$ is applied to the vertical lines connected to the electrodes of the memristors. For instance, to execute the clone operation on the first two devices in a row, $V_C$ is applied at $V_{c,1}$ while ground is applied at $V_{c,0}$. Gate voltages $V_{s,0}$ and $V_{s,1}$ are applied to open both transistors ($>$2V in this case). The first row ($V_{r,0}$) remains floating, facilitating cloning as illustrated in Fig.~\ref{fig:clone_explain}(b), effectively transferring data from the first device to the second. Similarly, voltages can be applied to rows, columns, and gates of a horizontal crossbar, as depicted in Fig.~\ref{fig:crossbars}(d), to execute the cloning operation. Fig.~\ref{fig:crossbars}(c) illustrates the implementation of the bit cloning operation in the horizontal crossbar. It is noteworthy that the operation conducted on the row-wise vertical crossbar resembles the operation in the horizontal crossbar for the column due to crossbar connections. The voltage sequence in various configurations enables data to be cloned bit-wise and column-wise, and even the entire word can be cloned to another word. Subsequently, we delve into the specifics of each operation. It is important to note that we exclusively focus on the vertical crossbar as similar operations can be performed on the horizontal crossbar.

\subsection{Bit Cloning}
In bit cloning, a single bit of data is moved either within the same row or column, although it can be transferred to any location within the same row or column. Row-wise and column-wise operations require distinct voltage schemes. A $3 \times 3$ crossbar structure elucidates the cloning methodology shown in Fig.~\ref{fig:methods}. The device in operation is highlighted using a solid line, while the dashed line indicates the unselected devices for that specific operation. Fig.~\ref{fig:methods}(a) presents the states of devices in a 3x3 crossbar, where `A' represents the LRS cell while all other cells are in HRS.

\subsubsection{Row-Wise}
In row-wise bit operation, `A' has to be moved within the same row, as depicted in Fig.~\ref{fig:methods}(b). The voltage applied to the device follows the configuration shown in Fig.~\ref{fig:crossbars}(b), selecting the first two devices connected in $r,0$. The gates of selected devices are connected to $V_{s,(0,1)} = V_g$ while other devices are deselected by applying 0V to their gates. During the cloning phase, $V_C$ is applied at $V_{c,1}$ and $V_{c,0}$ is connected to GND. The $r,0$ line remains floating, while other row lines are connected to a voltage, $V_{r,x} = V_{C}/2$, which is due to shared $V_{s,z}$, where $1\le x<m$ and $0\le z<2$. This results in proper device selection, which is shown in Fig.~\ref{fig:methods}(c), and clone `A' to the desired cell. It's important to note that $V_C$ is applied across the output device during the row-wise cloning phase. 

\subsubsection{Column-Wise}
Similar to row-wise bit cloning, column-wise bit cloning allows data to be cloned within the column, requiring a distinct voltage configuration in the opposite direction compared to row-wise bit cloning. Fig.~\ref{fig:methods}(d) displays the current state of devices after bit-wise cloning and the subsequent operation with the green arrow. Fig.~\ref{fig:methods}(e) shows the selected and unselected cells for operation. In column-wise cloning, a single gate line is shared among the devices in the column, with the commonly shared line being a row line rather than a column line, as in row-wise bit cloning. The required voltages are applied to the column side. To select the cell as shown in Fig.~\ref{fig:methods}(d), $V_{s,0}$ is connected to $V_g$, while all other gates are connected to 0V. During the cloning phase,  $V_{r,0}$ is connected to $V_{C}$, and $V_{r,1}$ is connected to the ground, while all other row voltages are connected to $V_{C}/2$. The column-wise operation requires $V_C$ at the input device, which creates a positive voltage drop at the device electrode, resulting in state change for the output device. The applied voltage terminal is opposite to the operation during row-wise cloning, where the $V_{C}$ is applied at the output device.

\begin{figure}[t]
    \centering
    \includegraphics[width=0.95\linewidth]{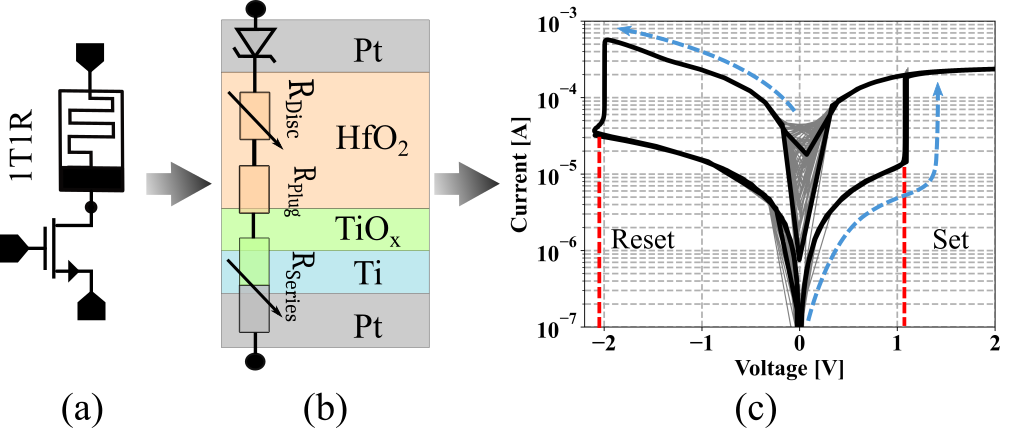}
    \caption{(a) 1T1R cell schematic, (b) material stacks of memristor, (c) I-V characteristics for 100 cycles}
    \label{fig:device_char}
\end{figure}

\begin{figure*}[ht!]
    \centering
    \includegraphics[width=0.95\linewidth]{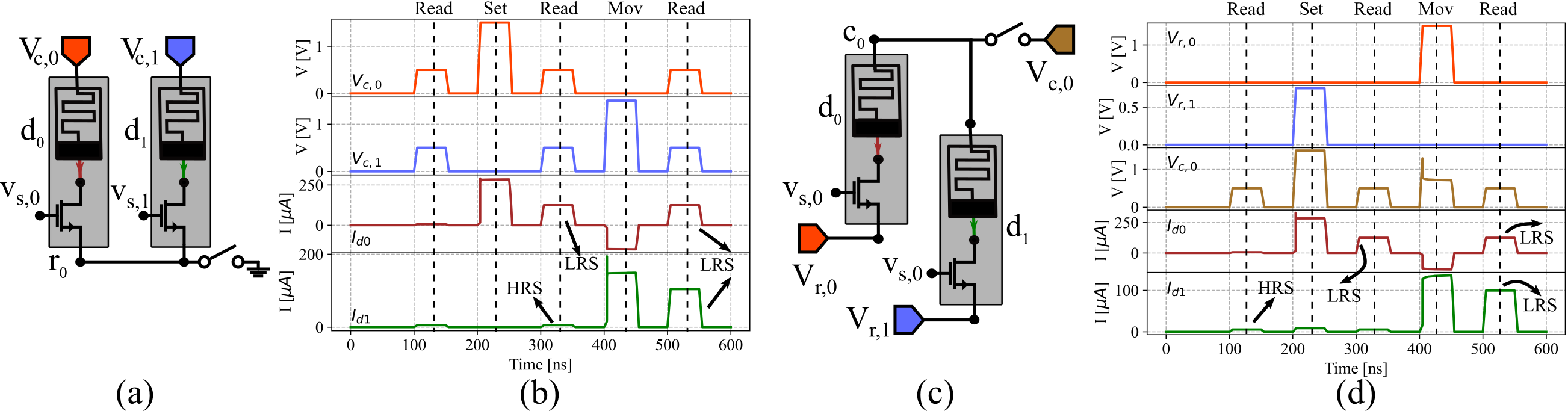}
    \caption{(a) Schematic for bit-wise cloning in the same row. In the schematic, we have four voltage sources, $V_{c,0}$, $V_{c,1}$, $V_{r,0}$, $V_{r,1}$, and currents in the memristors are marked in colors. (b) The simulation results' waveform shows each cycle marked on the top of the graphs. The last Read cycle shows the state of both devices. (c) Column-wise bit operation. The voltage sources are at $r_0$ and $r_1$ along with $c_0$ instead of $c_0$ and $c_1$. (d) Waveform during the column-wise bit operation.}
    \label{fig:row_col_results}
\end{figure*}

\subsection{Word Cloning}
As in a vertical crossbar, the gates are shared vertically (refer Fig.~\ref{fig:crossbars}(a)); it's possible to do the operation in parallel by opening the gate of all the lines connected vertically, allowing for the complete row to be cloned to another row.
As depicted in Fig.~\ref{fig:methods}(f), the example involves moving the first row to the last row (marked by the green arrow). The first Row contains two `A' cells. For word cloning operation, $V_{s,z}$ is connected to $V_g$, where $0\le z<n$, and $V_{r,0}$ is connected to $V_{C}$, $V_{r,1}$ is connected to ground, while $V_{c,x}$ remains floating, where $0\le x<n$. The unselected rows are connected to $V_{C}/2$. Fig.~\ref{fig:methods}(g) depicts the selected and unselected devices after applying the specified voltages across the row and column lines. This allows the complete row to be moved to another row, as each clone operation can be performed in parallel with a complexity of one cycle only. The operation is similar to the column-wise operation, where all column operations happen simultaneously. Fig.~\ref{fig:methods}(h) illustrates the final state after all operations.

All operations are depicted with respect to the vertical crossbar structure. Nevertheless, similar operations can also be conducted on the horizontal crossbar. Additionally, due to the crossbar structure, the operations interchange between row-wise and column-wise. Furthermore, in the vertical crossbar, word cloning conducted row-wise will transition to column-wise word cloning. The voltage required for the basic clone operation will remain the same.

\section{Results}
\label{results}
This section unveils the outcomes derived from SPICE level simulation using the Cadence spectre. The JART VCM v1b model has been used as an RRAM and the \textit{gpdk 45nm} technology node for CMOS integration. This section also provides insights into the energy consumption associated with cloning operations. 

\subsection{1T1R RRAM Switching Characteristics }
The 1T1R cell design comprises a single memristor and one NMOS transistor, forming the 1T1R structure. Fig.~\ref{fig:device_char}(a) illustrates the cell design alongside the equivalent stack of the memristor depicted in Fig.~\ref{fig:device_char}(b). The memristor model utilized in this investigation is experimentally validated and consists of a Pt/Ti/TiO$\rm_x$/HfO$\rm _2$/Pt material stack known for its favorable electroforming voltage and thermal stability. The I-V characteristics of the 1T1R cell are depicted in Fig.~\ref{fig:device_char}(c), which shows Set and Reset voltages marked with the red lines, with the Set voltage approximately 1V, and the Reset voltage approximately 2V. Any voltage between the Set and Reset voltages can be employed for the Read operation, with 0.5V utilized in this study to read the device state. Throughout the I-V characteristics, the gate of the transistors is maintained at~2.5V.
 
\subsection{Cloning Implementation}
The schematic depicting row-wise cloning in the vertical crossbar structure is presented in Fig.~\ref{fig:row_col_results}(a). Initially, all devices are set to HRS by default. To validate the cloning operation, $d_0$ is initialized to LRS/HRS by applying the necessary voltage at $V_{c,0}$. Fig.~\ref{fig:row_col_results}(b) has the waveform applied to $V_{c,0}$ and $V_{c,1}$, along with the current flowing through devices $d_{0}$ and $d_{1}$, denoted as $I_{d0}$ and $I_{d1}$, respectively. A Read pulse of 0.5V is initially applied across both devices to ascertain their current state. Subsequently, a Set voltage is applied to $V_{c,0}$ to transition device $d_{0}$ to LRS, as indicated in the subsequent read cycle. The `Mov' pulse marked in Fig.~\ref{fig:row_col_results}(b) signifies the clone operation pulse. During the clone operation, $V_{c,0} = 0V/GND$ and $V_{c,1} = V_C$ are applied, while $r_0$ remains floating. The final read operation verifies the state of both devices, confirming that both are in LRS, thus validating the clone operation.

Fig.~\ref{fig:row_col_results}(c) shows the simulated schematic for column-wise bit cloning. Similar to row-wise cloning, all devices are initially in HRS. However, the voltage scheme differs due to the crossbar connection. After Set and Read pulses, the clone operation is executed by applying $V_{r,0} = V_{C}$ while $V_{r,1} = 0V$. The column line $V_{c,0}$ remains floating during the operation. The final Read cycle confirms the successful cloning of $d_0$ data to $d_1$, as shown in Fig.~\ref{fig:row_col_results}(d).

\begin{figure}[t!]
    \centering
    \includegraphics[width=0.75\linewidth]{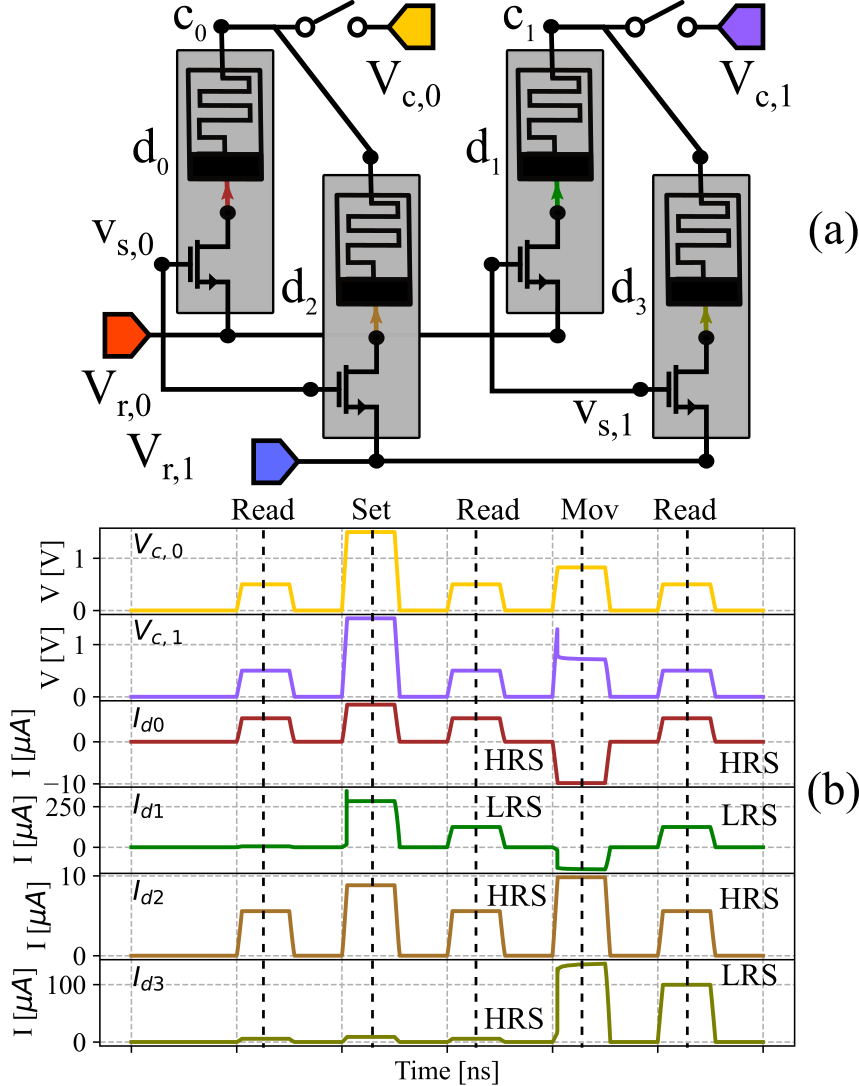}
    \caption{(a) Schematic for word cloning, showing 2x2 crossbar, where the data in $d_0$ and $d_1$ will be cloned to $d_2$ and $d_3$, respectively. The current flowing from each device is marked in colors. (b) Waveform for the cloning ``10"}
    \label{fig:word_cloning}
\end{figure}

Finally, word cloning on a 2x2 crossbar structure is depicted in Fig.~\ref{fig:word_cloning}. Similar to column-wise bit cloning, multiple columns share the same gate and row lines in the vertical crossbar, enabling parallel data cloning. As shown in Fig.~\ref{fig:word_cloning}(a), the same $V_{r,0}$ is applied to devices $d_{0}$ and $d_2$, while similarly, $V_{r,1}$ is applied to devices $d_1$ and $d_3$. Since $V_{s,0}$ and $V_{s,1}$ are shared for devices connected in $r_0$ and $r_1$, respectively, both column lines $c_0$ and $c_1$ are kept floating during the `Mov' cycle. Fig.~\ref{fig:word_cloning}(b) shows the waveforms for performing word cloning. Devices $d_0$ and $d_1$ are programmed to logic `1' and logic `0', respectively. The state of $d_0$ and $d_1$ after the clone cycle (Mov) can be observed in the final Read operation, reflecting the transfer of $d_0$ and $d_1$ states to $d_2$ and $d_3$, respectively.

\begin{table}[b]
    \centering
   \caption{Energy consumption for cloning}
    \label{tab:energy_init}
  \setlength{\tabcolsep}{4.5pt}       
    \begin{tabular}{|c|c|c||c|c|c|}
    \hline
    \multicolumn{3}{|c||}{Device operations} & \multicolumn{3}{c|}{Cloning} \\
    \hline
    \hline
        Operation & \makecell{Voltage \\ V}  & \makecell{Energy \\ (pJ)} &  Operation & \makecell{Voltage \\ V}  & \makecell{Energy \\ (pJ)}  \\
        \hline
        Reset & 2.25  & 15.54 & Bit (1) & 1.5   & 9.52  \\
        \hline
         Set & 1.5 & 20.17 &Bit (0) & 1.5 & 0.71  \\
        
         \hline
         Read (0 \& 1) & 0.5  & 3.1 & Word (2-bit) & 1.5   & 11.28  \\
         \hline
      
    \end{tabular}

\end{table}

\subsection{Energy and Latency Calculations}
Comparing energy consumption and latency of existing methods with the IMM approach is essential for assessing efficiency. Table~\ref{tab:energy_init} outlines energy usage during device operations such as Set, Reset, Read, and cloning, classified as bit and word cloning. The energy is obtained by multiplying the voltage waveforms with the sensed current and then integrating the product over the measurement time as per $\int_{0}^{t} v(t) \times i(t) \,dt$, where $t$ is the pulse time. In existing computing, copying operations typically require two cycles: read and writeback, consuming approximately 18 pJ and 23 pJ for copying logic `0' and logic `1', respectively, in the worst case. In terms of latency, the literature shows the two cycles for copying; in~\cite{bhattacharjee2020contra}, one for reading and another for writeback, and in~\cite{Padberg2024Experimental}, two NOT operations to copy. In contrast, the proposed IMM scheme enables data cloning within the memory in one cycle, requiring around 10 (9.52 + 0.71) pJ for bit cloning. An average 2-bit word cloning consumes only 11.28 pJ of energy. 
The proposed scheme reduces overall energy consumption by the sum of the energy expended during a read operation and the energy consumed by the peripheral during that operation. This reduction is particularly significant in large-scale applications. Additionally, it achieves a 2$\times$ improvement in latency.

\section{Discussion}
\label{discuss}

This study demonstrates a technique for cloning data within memory without reading. 
Table~\ref{tab:energy_init} shows that bit-wise (`0' \& `1') cloning consumes approximately 10.2 pJ, as the output cell is in a HRS, limiting switching current and reducing energy consumption. For 2-bit word cloning, the average energy is dominated by the ``11" combination at around 22 pJ, while ``00" consumes 0.7 pJ, and ``01 \& 10" consumes 11.11 pJ. These energy values are based on the operational voltage, which can be adjusted for optimization. However, it is assumed that all devices are in an HRS, similar to the logic operation implementation. If the cell has been previously utilized, it must be switched to HRS before the cloning operation, necessitating an additional cycle. This procedure is similar to handling a reused cell and initializing the output cell to HRS. For output devices in use or unknown states, dependencies can be verified at the compiler level and initialized with the output device for logic operations, like regular memory write operations. Since the voltage requirements are compatible with other logic operations, the area needed for the control circuit for cloning will remain the same. 


\section*{Conclusion}
\label{conc}
This paper introduced the concept of IMM in RRAM memory crossbars, enabling data cloning without reading the state within a single cycle. IMM facilitates efficient data cloning within the memory, eliminating the necessity for read and write-back cycles and resulting in substantial energy savings. Moreover, the scheme supports bit-wise data movement in both columns and rows while enabling parallel cloning of entire columns with the complexity of a single cycle. The effectiveness of the IMM concept has been demonstrated through SPICE-level simulation analysis with an experimentally validated RRAM model. Furthermore, we have thoroughly examined the energy consumption and latency associated with our proposed technique. The IMM scheme exhibits energy efficiency and is  2$\times$ faster than the existing method of copying data in the RRAM crossbar. Moving forward, we plan to validate the proposed scheme experimentally using a fabricated RRAM crossbar.

\section*{Acknowledgments}
This work was supported in part by the Federal Ministry of Education and Research (BMBF, Germany) in the project NEUROTEC II under Project 16ME0398K, Project 16ME0399, German Research Foundation (DFG) within the Project PLiM (DR 287/35-2) and through Dr. Suhas Pai Donation Fund at IIT~Bombay.

\bibliographystyle{IEEEtran}
\bibliography{Bib}

\end{document}